\documentclass[twocolumn,aps,prb,showpacs,amssymb]{revtex4}

\usepackage{graphicx}
\usepackage{dcolumn}
\usepackage{bm}

\newcommand{\ep}{\varepsilon}
\renewcommand{\k}{{\bf k}}
\def\bsigma{\mbox{\boldmath$\sigma$}}

\begin{document}

\preprint{}

\title{Energy relaxation due to magnetic impurities in mesoscopic
  wires:\\ Logarithmic approach}

\author{O. \'Ujs\'aghy$^a$}
\author{A. Jakov\'ac$^{a}$}
\author{A. Zawadowski$^{a,b}$}
\affiliation{$^{a}$Budapest University of Technology and Economics,
Institute of Physics and Research group Theory of Condensed Matter of 
Hungarian Academy of Sciences
H-1521 Budapest, Hungary}
\affiliation{$^{b}$Research Institute for Solid State Physics, POB 49, H-1525
Budapest, Hungary}

\date{\today}

\begin{abstract}
  The transport in mesoscopic wires with large applied bias voltage
  has recently attracted great interest by measuring the energy
  distribution of the electrons at a given point of the wire, in
  Saclay. In the diffusive limit with negligible energy relaxation
  that shows two sharp steps at the Fermi energies of the two
  contacts. Those steps are, however, broadened due to the energy
  relaxation which is relatively weak if only Coulomb
  electron-electron interaction is assumed. In some of the experiments
  the broadening is more essential reflecting an anomalous energy
  relaxation rate proportional to $E^{-2}$ instead of $E^{-3/2}$ where
  $E$ is the energy transfer. Later it has been suggested that such
  relaxation rate can be due to electron-electron interaction mediated
  by Kondo impurity which is, as it is known since a long time,
  singular in $E$. In the present paper the magnetic impurity mediated
  interaction is systematically studied in the logarithmic
  approximation valid above the Kondo temperature. In the case of
  large applied bias voltage Kondo resonances are formed at the steps
  of the distribution function and they are narrowed by increasing the
  bias. An additional Korringa energy broadening occurs for the spins
  by creating electron-hole pairs in the electron gas out of
  equilibrium. That smears the Kondo resonances, and the renormalized
  coupling can be replaced by a smooth but essentially enhanced
  average coupling and that enhancement can reach the value 8-10. Thus
  the experimental data can be described by formulas without
  logarithmic Kondo corrections, but with enhanced coupling. In
  certain regions of large bias, that averaged coupling depends weakly
  on the bias. In those cases the distribution function depends only
  on the ratio of the electron energy and the bias, showing scaling
  behavior. The impurity concentrations estimated from those
  experiments and other dephasing experiments can be very different.
  An earlier paper has been devoted to provide a possible explanation
  for that by considering the surface spin anisotropy due to strong
  spin-orbit interaction in the host which can hinder the motion of
  spins with integer spin value.
\end{abstract}

\pacs{73.23.-b,72.15.Qm,71.70.Ej}

\maketitle 

\section{Introduction}
In the last few years the study of inelastic electron scattering mechanisms in
mesoscopic metallic systems has attracted considerable interest both
experimentally and theoretically.  The motivation is to identify the
mechanisms responsible for destroying the quantum coherence of the
quasiparticles.  In one kind of the experiments the dephasing time is
determined from measurements of magnetoresistance
\cite{magexp1,magexp2,Natelson,AHexp,Schopfer}, universal conductance
fluctuation \cite{Mohanty}, and Aharonov-Bohm oscillation in mesoscopic rings
in magnetic fields \cite{AHexp}. In the other kind of experiments on short
wires with large bias voltage the non-equilibrium distribution functions of
the electron energies are measured \cite{Pothier,PothierAg,Pothier2} which
provides an indirect information about the energy relaxation of the electrons.
The dephasing and energy relaxation are due to inelastic scattering processes,
thus we can obtain information on those from these two kinds of experiments.
The experiments were performed at very low temperatures (below about $1$K)
where the electron-phonon scattering is weak, thus the main inelastic
processes are the electron-electron scattering and the scatterings by
dynamical defects (e.g.  magnetic impurities, two-level systems).

In this paper we investigate theoretically the energy relaxation due to
magnetic impurities in thin metallic wires with large bias voltage. Our
motivation is that there are many cases where deviations from the expectations
of the theory of Coulomb electron-electron interaction by Altshuler, Aronov
and coworkers \cite{Alt} were found. The Saclay group \cite{Pothier} analyzing
the experimental data suggested a phenomenological anomalous effective energy
relaxation rate proportional to $E^{-2}$ where $E$ is the energy transfer.
That differs from the one due to the Coulomb interaction which varies like
$E^{-3/2}$. That anomalous relaxation rate was attributed to the presence of
magnetic impurities either implanted or contained by the starting material as
contamination or dynamical defects \cite{KG,th1,th2,th3,th4,GGA1}.

The main goal of our earlier paper \cite{UJZ} has been to suggest a
resolution of the discrepancy concerning the required magnetic
impurity concentrations to fit the different experimental data and in
some cases those discrepancies may reach the level of two orders of
magnitude. Therefore, even the role of magnetic impurities was
questioned in spite of the observed sensitivity on external magnetic
field \cite{Anthore}. Having a thin wire the magnetic surface
anisotropy \cite{UZ1} has been taken into account in
Ref.~\onlinecite{UJZ} which hinders the motion of the spin and in that
way reduces the Kondo scattering. The previous theoretical works
\cite{KG,th1,th2,th3,th4,GGA1}, however, showed the phenomena is
rather complex and it is very hard to separate and identify the role
of the different effects. That is especially valid for the
non-crossing approximation which is based on extensive numerical work
\cite{th2,th4} and many details are hidden.  The present paper is
devoted to those problems, and the role of surface anisotropy has
already been discussed in Ref.~\onlinecite{UJZ}.

Considering the Kondo effect the different temperature and energy ranges have
very different characters. Below the characteristic Kondo energy the Fermi
liquid behavior is recovered. That range has been intensively studied even in
the non-equilibrium situation in case of wires, but the situation in quantum
dots shows strong similarities \cite{Coleman,Oguri}. The most difficult range
is around the Kondo energy, while well above that perturbative approaches are
appropriate, known as logarithmic approximation. The main idea is to apply
perturbation theory and to collect the terms with the highest power of the
characteristic logarithm in each order. That leading logarithmic approach is
well established. It will be shown that most of the experiments fall in that
range.

In case of non-equilibrium situation the voltage applied on the short
wire plays crucial role which has to be incorporated into that
scheme. 

The applied voltage $U$ occurs in two different ways: 

(i) \emph{Reduction of the Kondo temperature}: In a short wire in the
diffusive limit with a weak energy relaxation the electron comes from one of
the two contacts without energy loss. The distribution of those electrons is
characterized by different Fermi energies which are split by the applied
voltage, $eU$ where $e$ is the electronic charge. The Kondo effect arises from
those regions where there are sharp changes in the energy distribution
functions, thus at the two Fermi energies. At both places separate Kondo
resonances develop, but the amplitudes of sharp changes in the distribution at
both energies are reduced compared to the equilibrium case, where the change
is the sum of those two.  Therefore, in the wire the equilibrium Kondo
temperature for the step with e.g. the lower energy is reduced at position $x$
as \cite{th3,GGA1,Coleman2}
\begin{equation}
  \label{reducedTK}
T_K(x) = \frac{T_K ^{\frac{1}{1-x}}}{(eU)^{\frac{x}{1-x}}}  
\end{equation}
which is valid for $T_K<eU$ where $x$ is measured in the units of the length
of the wire $L$. The voltage $U$ measures the energy range where the
non-equilibrium electron distribution takes place.  That reduction in $T_K$
can be very essential, e.g. for $T_K=0.3\,$K, $eU=0.3\,$meV$\sim 3\,$K, and
$x=0.5$, $T_K(x=0.5)=0.1T_K=0.03\,$K. Thus for a considerable voltage and for
Kondo temperature in a typical range ($T_K=0.1-1\,$K) the reduced Kondo
temperature is rather small compared to the energy scale $eU$ playing role in
the electronic transport. That ensures the applicability of the logarithmic
approach as far as $T_K\ll eU$. In other cases, e.g.  the non-crossing
approximation must be applied \cite{th2,th4}.

(ii) \emph{Enhanced Korringa relaxation rate}: the Korringa relaxation
\cite{Korringa} describes the impurity spin relaxation due to electron-hole
creations. The creation of those with almost zero energy can be only in the
energy range where the sharp single-step ($T=0$) electron distribution is
modified. That can be achieved by raising the temperature or applying the
voltage. In the latter case the extra relaxation rate is
\begin{equation}
  \frac{\hbar}{2\tau_K} = 2\pi S(S+1) (\rho_0 J)^2 x(1-x) e U
\label{Korr_eq}
\end{equation}
at $T=0$ where $S$ is the spin of the impurity and $\rho_0 J$ is the
dimensionless electron-impurity spin coupling. In the equilibrium case
the Korringa relaxation rate is not part of the logarithmic approach,
but $eU\gg T_K$ makes it non-negligible.

Namely, in the next to leading logarithmic approximation the real part
of the pseudospin particle self-energy contains a single log which is
even lost in the imaginary part. That log is not contained by the
simple logarithmic approach.  The effect of the imaginary part (the
inverse lifetime of the spin) remains, however, very crucial. For
spins there are different relaxation rates e.g. the $T^{-1}_1$ and
$T^{-1}_2$ in the ESR, NMR experiments. The actual values are
consequences of a delicate balance between the self-energy and vertex
corrections, which can modify the prefactors like $S\,(S+1)$ in
Eq.~(\ref{Korr_eq}) \cite{Walker}. Cancellation of that type are very
well known in the scaling equation of the renormalized effective
coupling, where the value of the spin $S$ drops out \cite{FZ-AM}.
Considering the paramagnetic impurity electron-spin resonance that has
been discussed earlier in detail \cite{Walker} and it is found that in
the $T^{-1}_2$ relaxation rate the factor $(S+1)$ drops out.  Thus the
prefactor $S\,(S+1)$ in Eq.~(\ref{Korr_eq}) may result in an essential
overestimation for $S\ge 1$. A systematic study of that problem is not
known in the present non-equilibrium situation \cite{WolfleKroha}.
Therefore, the consequences of the present ambiguity is only
demonstrated by introducing a multiplying factor $\lambda$ of order of
unity ($1/2\le\lambda\le 2$) for the relaxation rate given by
Eq.~(\ref{Korr_eq}) and the role of $\lambda$ will be discussed. It is
reasonable to assume, that the factor $\lambda$ imitating those
cancellations is more important for larger spins $S>1$.

In the following the logarithmic approach is applied for the
non-equilibrium case, and the anomalous Korringa relaxation is
additionally built in.

Considering the theoretical methods two ways can be followed: (i) The Keldysh
Green function technique devoted to the non-equilibrium \cite{th2,th4,Coleman}
which may be combined with the non-crossing approximation \cite{Ueda}. (ii)
Direct calculation of the scattering rate by the time ordered perturbation
theory starting with an arbitrary state and taking into account the actual
distributions in the occupations of the intermediate and final states
\cite{th3,Paaske}. Restricting to logarithmic accuracy the second method
is very easy to apply, in other cases one should turn to the first one.

The paper is organized as follows. In Section \ref{sec:2} the
transport phenomenon is imposed. In Section \ref{sec:3} the kernel in
transport equation due to the magnetic impurity mediated
electron-electron scattering is determined using the leading
logarithmic approximation. The electron distribution and exchange
coupling renormalization is carried out in a self-consistent way. In
Section \ref{sec:4} the Korringa relaxation is taken into account in
the kernel and in the smearing of the renormalized coupling, and then
it is also determined in a self-consistent way with the other
quantities. In Section \ref{sec:5} the results are analyzed from the
point of view of the importance of the different ingredients of the
theory. In the Conclusion the results obtained are summarized also
from the point of view of Ref.~\onlinecite{UJZ}, where the simplified
method is applied for the case where the magnetic surface anisotropy
\cite{UZ1} is included.

\section{Non-equilibrium electron energy distribution in diffusive wires}
\label{sec:2}

\begin{figure}
\begin{center}
\includegraphics[height=1.8cm]{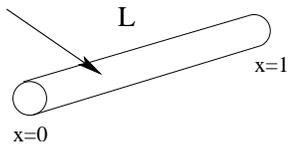}
\caption{The wire of length L. The arrow points to the position $x$ of the
attached tunnel junction which is measured in units of $L$.}
\label{fwire}
\end{center}
\end{figure}

In the experiments the distribution of electron energies in a wire with large
bias was measured at different position by attaching extra tunnel junction to
the side of the wire (see Fig.~\ref{fwire}). According to that we consider a
wire with length $L$ and the distribution function at position $x$ in the wire
in units of $L$ for energy $\ep$ and at time $t$ is denoted by $f(\ep,x,t)$.
The non-equilibrium distribution function in
the diffusive limit is determined by the time dependent Boltzmann equation 
\begin{equation}
\label{BE}
    \frac{\partial f(\ep,x,t)}{\partial t}-\frac{1}{\tau_D}
      \frac{\partial^2 f(\ep,x,t)}{\partial^2 x} + I_{\mathrm
        coll.}(\{f\})=0
\end{equation}
where $\tau_D=\frac{L^2}{D}$ is the diffusion time and $D$ is the diffusion
constant. In the followings the stationary solution is taken, thus the time
dependence is dropped and $f$ is assumed to be independent of the spin. The
collision integral $I_{\mathrm coll.}(\{f\})$ due to inelastic scatterings in
Eq.~(\ref{BE}) can be expressed by the scattering rate $W(\ep, E)$ of
electrons of energy $\ep$ with energy transfer $E$ as
\begin{eqnarray}
\label{Icoll}
    I_{\mathrm coll.}(\{f\})&=&\int dE
   \bigl \{f(\ep) [1-f(\ep-E)] W(\ep, E)\nonumber\\
    &-&[1-f(\ep)] f(\ep-E) W(\ep-E,-E) \bigr \}.
\end{eqnarray}
In absence of inelastic scattering processes $I_{\mathrm coll.}(\{f\})=0$,
thus the solution of Eq.~(\ref{BE}) is a double-step distribution
function \cite{Boltzmann-eq}
\begin{equation}
  \label{eq:freesol}
  f^{(0)}(\ep,x)=(1-x) n_F (\ep-\frac{e U}{2})+x n_F (\ep+\frac{e U}{2})
\end{equation}
where $U$ is the voltage between the ends of the wire.
The sharp steps are smeared due to the inelastic processes even at very low
temperature. Taking into account inelastic scattering in $W$ and starting with
the solution Eq.~(\ref{eq:freesol}), the Boltzmann
equation can be solved self-consistently at least numerically.

In the numerical calculation searching for the stationary solution the
Boltzmann equation is solved by iteration. In the iteration we start with the
solution in absence of inelastic processes (Eq.~(\ref{eq:freesol})) and after
at each step of the iteration the collision integral is calculated from the
actual distribution function and scattering rate and then the new distribution
function satisfying the boundary conditions is determined. The variables $x$
and $\ep$ are discretized usually in 20-60 points. The number of the necessary
iterations depends on the approximations used in the calculation of the
collision integral, but usually the final solution is reached by 20-120
iteration.

For electron-electron interaction the scattering rate $W$ can be expressed as
\begin{equation}
\label{W-kernel}
W(\ep,E)=\int d\ep'K(E,\ep,\ep') f(\ep') [1-f(\ep'+E)]
\end{equation}
where the kernel $K(E,\ep,\ep')$ is determined by the specific interaction
and $\ep'$ is the energy after the collision.
 
The predicted dependence of the kernel $K$ on $E$ in case of the Coulomb
electron-electron scattering \cite{Alt} is $\sim E^{-3/2}$ in $1D$ regime
which explains perfectly some of the experiments \cite{PothierAg}, but in
other cases other relaxation mechanisms must exist. The Saclay group
\cite{Pothier} introduced phenomenologically an extra interaction between the
electrons with $K\sim 1/E^2$ singularity.  It is known since a long time
\cite{SZ} that the interaction mediated by magnetic impurities is singular in
the energy transfer and recently Kaminski and Glazman \cite{KG} called the
attention to similar $1/E^2$ singularity in the magnetic impurity mediated
electron-electron interaction kernel in the leading order of perturbation in
the Kondo coupling.  At that time authors of
Ref.~[\onlinecite{KG,th1,th2,th3,th4,GGA1}] suggested that the energy exchange
is mediated by Kondo impurities (magnetic and structural defects).

In the following we perform a systematic study of the Boltzmann
equation (\ref{BE}) for electron-electron interaction mediated by Kondo
impurities and the different approximations are built in step by step.

Assuming low concentration of the impurities we will use the single impurity
Kondo model \cite{Kondo}
\begin{equation}
\label{Kondomodel}
  H= \sum\limits_{\k,\sigma} {\ep_{\k}}
    a_{\k\sigma}^\dagger a_{\k\sigma} +J_0
    \sum\limits_{{\k\k'}\atop{\sigma\sigma'}} \mathbf{S}\,
    a_{\k\sigma}^\dagger \bsigma_{\sigma\sigma'} a_{\k'\sigma'}
\end{equation}
where $a_{\k\sigma}^\dagger$ creates a conduction electron with momentum $\k$,
spin $\sigma$ and energy $\ep_{\k}$ measured from the Fermi level. The
conduction electron band is taken with constant energy density $\rho_0$ for
one spin direction, with a sharp and symmetric bandwidth cutoff $D_0$,
$\bsigma$ stands for the Pauli matrices, $J_0$ is the Kondo coupling, and
$\mathbf{S}$ is the spin operator of the impurity.

\begin{figure}
\centerline{\includegraphics[height=1.5cm]{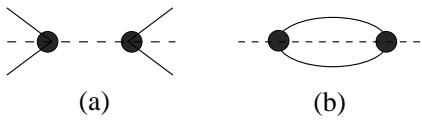}}
\caption{The diagrams used for calculating (a) the kernel and (b)
  the Korringa lifetime of the impurity spin. The
  solid lines denote the conduction electrons, the dotted lines the impurity
  spin, and the blob is the Kondo coupling.}
\label{fig2}
\end{figure}

From perturbation theory in the coupling $J_0$ the first non-vanishing
contribution to the kernel comes from the diagram in Fig.~\ref{fig2}(a) as
\cite{KG,th1}
\begin{equation}
  \label{pertkernel}
  K(E,\ep,\ep')=\frac{\pi}{2\hbar}\frac{c}{\rho_0} S (S+1) (2\rho_0 J_0)^4
  \frac{1}{E^2} 
\end{equation}
where $c$ is the concentration of the homogeneously distributed magnetic
impurities in the wire. That kernel gives back the
phenomenologically introduced energy dependence \cite{Pothier}, but its
strength may be too small \cite{Pothier2} according to the small bare Kondo
coupling $J_0$.

\section{Leading logarithmic approximation}
\label{sec:3}

To improve the calculation as the first step the renormalization of the
coupling in the presence of the applied voltage is taken into account. The
modified kernel reads
\begin{eqnarray}
  \label{runkernel}
 && 
  K(E,\ep,\ep')=\frac{2\pi c}{\hbar} \rho_0^3 S
 (S+1)\times\nonumber \\ 
 &&\times\biggl \{\frac{2}{E^2}\biggl ( S (S+1)\bigl [J(\ep) J(\ep'+E)-J(\ep')
 J(\ep-E)\bigr ]^2 \nonumber \\
 &&\hspace*{1em}+2 J(\ep) J(\ep') J(\ep'+E) J(\ep-E)\biggr )\nonumber \\
  &&-\frac{S (S+1)-1}{E (\ep-\ep'-E)}\biggl ([J^2(\ep)+J^2(\ep')] J(\ep'+E)
 J(\ep-E) \nonumber \\ 
  &&\hspace*{3em}-[J^2(\ep-E)+J^2(\ep'+E)] J(\ep) J(\ep')\biggr ) \biggr \}
\end{eqnarray} 
where $J(\ep)$ is the renormalized coupling and the approximations are made in
spirit of logarithmic approximation discussed in detail in the Appendix.
Considering only the most divergent terms proportional to $1/E^2$ in the limit
$E\rightarrow 0$ and ignore the $E$-dependence in the arguments of $J$ in
Eq.~(\ref{runkernel}) we get back the result of Eq. (14) of Ref. \cite{th1}
thus the kernel is proportional to $J^2(\ep) J^2(\ep')/E^2$. The agreement
with the result Eq.~(10) of Ref.~[\onlinecite{KG}] is also only after taking
the limit $E\rightarrow 0$ in $J$. In general, however, this factorization is
not true, the kernel contains mixed terms as well.

In our non-equilibrium wire case the renormalized Kondo coupling
depends also on the position $x$ in the wire. In the leading
logarithmic approximation carrying out similar resummation as in the
equilibrium case, the leading logarithmic scaling equation is
  \begin{equation}
\label{LSE}
    \frac{\partial J(\ep,x)}{\partial (\ln\frac{D_0}{D})}=2 \rho_0 J^2 
(f(\ep-D)-f(\ep+D))
  \end{equation}
  where the original bandwidth cutoff $D_0$ is reduced to $D$. In
  Eq.~(\ref{LSE}) $|\ep|\ll D$ is assumed, thus at the start the electron band
  can be taken always symmetric to $\ep$, which assumption is not applicable
  in general. On the right hand side the actual distribution functions occur
  which must be determined in a self-consistent way.  Assuming a single step
  like distribution that gives back the well known result \cite{Korringa} and
  with a double step the ones in Ref.~[\onlinecite{th3}] and
  Ref.~[\onlinecite{GGA1}].

The renormalized Kondo coupling in the leading
logarithmic approximation is the solution of Eq.~(\ref{LSE}):
  \begin{equation}
    \label{eq:JWK}
    J (\ep,x)=\frac{J_0}{1-2\rho_0
    J_0\int\limits_0^{D_0}\frac{d D'}{D'}(f(\ep-D')-f(\ep+D'))}.
  \end{equation}
  
\begin{figure}
  \begin{picture}(100,600)(100,100)
    \put(-150,-70){\includegraphics[height=30cm]{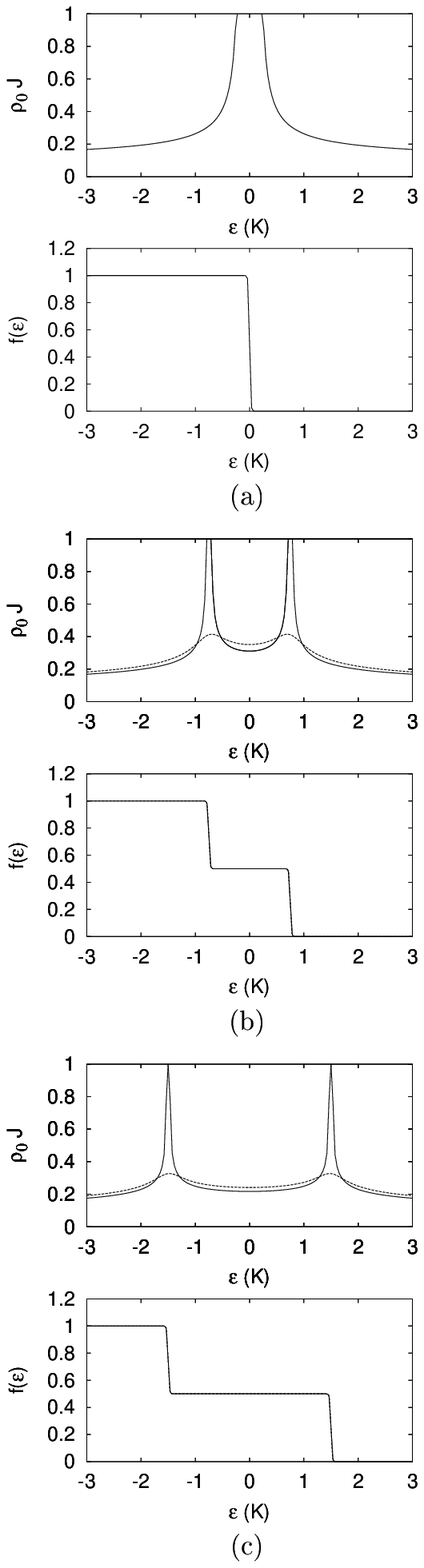}}
  \end{picture}
\caption{The renormalized coupling (top) calculated from the sharp double-step 
  distribution function (bottom) at $x=0.5$; $T=0.01$K, $T_K=0.15$K ($\rho_0
  J_0=0.039$, $D_0=5\cdot 10^4$K): (a) $U=0$mV, $\tau_K=\infty$ (b)
  $U=0.15$mV, solid line: $\tau_K=\infty$, dotted line:
  $\frac{\hbar}{2\tau_K}=0.25$K (c) $U=0.3$mV, solid line:
  $\tau_K=\infty$, dotted line: $\frac{\hbar}{2\tau_K}=0.25$K.}
\label{fig3a}
\end{figure}

\begin{figure}
  \begin{picture}(100,600)(100,100)
    \put(-150,-70){\includegraphics[height=30cm]{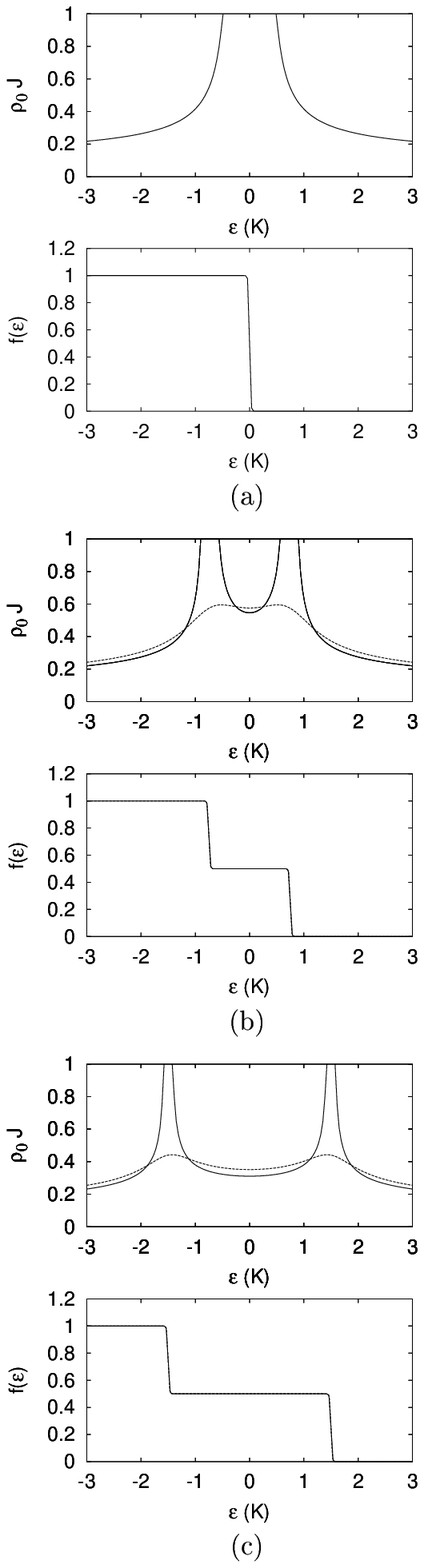}}
  \end{picture}
\caption{The renormalized coupling (top) calculated from the sharp double-step 
  distribution function (bottom) at $x=0.5$; $T=0.01$K, $T_K=0.3$K ($\rho_0
  J_0=0.042$, $D_0=5\cdot 10^4$K): (a) $U=0$mV, $\tau_K=\infty$ (b)
  $U=0.15$mV, solid line: $\tau_K=\infty$, dotted line:
  $\frac{\hbar}{2\tau_K}=0.4$K (c) $U=0.3$mV, solid line:
  $\tau_K=\infty$, dotted line: $\frac{\hbar}{2\tau_K}=0.4$K.}
\label{fig3b}
\end{figure}

  It is important to note that due to the finite voltage the width of
  the renormalized, resonant coupling $J(\ep)$ is essentially reduced
  compared to the equilibrium values (see Fig.s~\ref{fig3a} and
  \ref{fig3b}), thus the logarithmic approach can be valid even
  well below the equilibrium Kondo temperature $T_K$ for large bias
  $U$. Due to the finite voltage separate Kondo resonances are formed
  \cite{th2,th3,th4,GGA1} at the two Fermi energies corresponding to
  the two contacts thus $J(\ep,x)$ has two peaks (see Fig.s~\ref{fig3a}
  and \ref{fig3b}). The validity of the logarithmic approach is
  restricted by the condition $\rho_o J(\ep,x) < 1$ nearby the two
  Fermi energies which was always checked in the calculation by
  plotting the actual Kondo coupling with respect to the energy.

In case of $e\,U>T_K$ the Kondo temperature is reduced and the singularity is
smeared by the Korringa rate, thus the condition $\rho_0 J <1$ is satisfied in
most of the cases (see Fig.s~\ref{fig3a} and \ref{fig3b}).

In the numerical calculation $\rho_0 J(\ep,x) > 1$ is replaced by $\rho_0
J(\ep,x)=1$ as $\rho_0 J(\ep,x) > 1$ is always an artifact of the
approximation.

\section{Korringa relaxation rate}
\label{sec:4}
 
In the leading logarithmic order the self-energy corrections (wave function
renormalization and Korringa lifetime \cite{Korringa}) are neglected as they
appear only in higher order. At a finite voltage $U$, however, that inverse
lifetime proportional to $U$ results in anomalous broadening which can
be very essential. Therefore, the effect of $U$ is taken into account in the
spin spectral function $\rho_S (\ep)$ by an energy independent Korringa
lifetime $\tau_K$ as
\begin{equation}
 \label{rhops}
 \rho_{S}(\ep)=\frac{1}{\pi}\frac{\frac{\hbar}{2\tau_K}}{\ep^2
 +\frac{\hbar^2}{4\tau_K^2}}. 
\end{equation}
The energy dependence in $\tau_K$ leads to changes only in the tail of the
spectral function which is independent of $U$ for $|\omega| > eU$ and
therefore it is neglected in the crude leading logarithmic approximation
applied. The leading term of the Korringa relaxation rate is calculated
according to the self-energy diagram in Fig.~\ref{fig2}(b) as
\begin{equation}
  \label{tauK}
\frac{1}{2\tau_K(x)}=\frac{2\pi}{\hbar} \rho_0^2 S (S+1) 
\int d\ep J_K^2(\ep,x) (1-f(\ep,x)) f(\ep,x)
\end{equation}
where in the self-consistent calculation the Korringa modified couplings $J_K$
are used. In this way the effect of $U$ and the position dependence are taken
into account.
 
To include the effect of Korringa broadening the calculation of
$K(E,\ep,\ep')$ given by Eq.~(\ref{runkernel}) must be repeated using the
spectral function Eq.~(\ref{rhops}) resulting in the following form of the
kernel:
\begin{eqnarray}
  \label{korrkernel}
&&\hspace*{-1em}
K(E,\ep,\ep',\tau_K)=\frac{2\pi c}{\hbar} \rho_0^3 S (S+1)\times\nonumber \\
&&\hspace*{-0.5em}\times\biggl \{\frac{2 E^2}{(E^2+\frac{\hbar^2}{4\tau_K^2})^2}
\cdot\nonumber \\
&&\hspace*{1em}\cdot\biggl ( S (S+1)\bigl [J_K(\ep) J_K(\ep'+E)-J_K(\ep')
 J_K(\ep-E)\bigr ]^2 \nonumber \\
&&\hspace*{2em}+2 J_K(\ep) J_K(\ep') J_K(\ep'+E) J_K(\ep-E)\biggr )\nonumber \\
&&\hspace*{0.5em}+\frac{2 \frac{\hbar^2}{4\tau_K^2}}
{(E^2+\frac{\hbar^2}{4\tau_K^2})^2}\cdot\nonumber \\
&&\hspace*{1em}\cdot\biggl ( S (S+1)\bigl [J_K(\ep) J_K(\ep'+E)+J_K(\ep')
 J_K(\ep-E)\bigr ]^2 \nonumber \\
&&\hspace*{2em}-2 J_K(\ep) J_K(\ep') J_K(\ep'+E) J_K(\ep-E)\biggr )
\nonumber \\
&&\hspace*{0.5em}+\frac{(S (S+1)-1) E (\ep'-\ep+E)}{(E^2+\frac{\hbar^2}
{4\tau_K^2})
((\ep'-\ep+E)^2+\frac{\hbar^2}{4\tau_K^2})}\cdot\nonumber \\
&&\hspace*{1.5em}\cdot\biggl ([J_K^2(\ep)+J_K^2(\ep')] J_K(\ep'+E)
 J_K(\ep-E) \nonumber \\ 
&&\hspace*{2.2em}-[J_K^2(\ep-E)+J_K^2(\ep'+E)] J_K(\ep) J_K(\ep')\biggr )
\nonumber \\
&&\hspace*{0.5em}-\frac{(S (S+1)-1) \frac{\hbar^2}{4\tau_K^2}}
{(E^2+\frac{\hbar^2}{4\tau_K^2})
((\ep'-\ep+E)^2+\frac{\hbar^2}{4\tau_K^2})}\cdot\nonumber \\
&&\hspace*{1.5em}\cdot\biggl ([J_K^2(\ep)+J_K^2(\ep')] J_K(\ep'+E)
 J_K(\ep-E) \nonumber \\ 
  &&\hspace*{2.2em}+[J_K^2(\ep-E)+J_K^2(\ep'+E)] J_K(\ep) J_K(\ep')\biggr ) 
\biggr \}
\end{eqnarray}
where for the sake of simplicity we suppressed in the notation the spatial
dependence of the kernel and the Kondo coupling.
In Eq.~(\ref{tauK}) and (\ref{korrkernel})
$J_K(\ep)$ is the solution of the leading logarithmic scaling
equation modified by the finite Korringa lifetime
\begin{equation}
\label{LSE_K}
    \frac{\partial J_K(\ep)}{\partial (\ln\frac{D_0}{D})}=2 \rho_0 J_K^2(\ep)
\hspace*{-0.5em}\int\limits_{-\infty}^{\infty}\hspace*{-0.5em}
d\ep'\rho_{S}(\ep-\ep')\bigl (f(\ep'-D)-f(\ep'+D)\bigl ).
  \end{equation}
  The solution of Eq.~(\ref{LSE_K}) is the solution of the scaling equation
  Eq.~(\ref{eq:JWK}) with infinite spin lifetime smeared
  by the spin spectral function as
\begin{equation}
    \label{eq:J_K}
J_K^{-1}(\ep)=\!\int\limits_{-\infty}^{\infty}\! d\ep'
  \rho_{s}(\ep-\ep') J^{-1}(\ep').
  \end{equation}
  
  Thus at each step of the iteration solving the Boltzmann equation the
  smeared, renormalized Kondo coupling $J_K(\ep,x)$, the distribution function
  $f(\ep,x)$, and the Korringa lifetime $\tau_K(x)$ were updated.
  
  The spectral function given by Eq.~(\ref{rhops}) has further effects. In the
  most of the previous calculations the Korringa broadening is not taken into
  account (see e.g. Ref.~[\onlinecite{WolfleKroha}]) in the initial and
  final states. It is easy to show that including those broadenings in the
  collision term can be taken into account as an additional broadening in the
  singular part of the kernel. As it has already been discussed in the
  Introduction the vertex corrections should reduce that broadening. Thus the
  broadening is intensified in that way, but after the reduction by the vertex
  corrections that cannot be very different from the overestimated one given
  by Eq.~(\ref{tauK}). At the end the amplitude of the relaxation rate is
  given by Eq.~(\ref{tauK}) and that modification is taken into account by a
  phenomenological factor $\lambda$ ($1/2\le\lambda\le 2$) and its influence
  on the results will be discussed later.

\section{Results and discussion of the approximations}
\label{sec:5}

In the numerical calculation at each step of the iteration the
position dependent collision integral Eq.~(\ref{Icoll}) is calculated
in terms of the actual distribution function and the kernel defined in
Eq.~(\ref{W-kernel}) in the following way: first the renormalized
coupling $J$ is calculated from the actual $f$ smeared
(Eq.~(\ref{eq:J_K})) or not (Eq.~(\ref{eq:JWK})) by the Korringa
relaxation rate calculated from the preceeding $f$ and $J$.  Then the
new Korringa relaxation rate and after the kernel in terms of $\tau_K$
and $J$ are calculated according to the Eq.s~(\ref{tauK}) and
(\ref{korrkernel}). At the end of the given iteration step the new
distribution function is determined from the old distribution function
and the collision integral. The boundary conditions are satisfied at
each step of the iteration by construction.

In the followings the results are presented and the importance of the
different ingredients of the theory are analyzed.

\subsection{Renormalized couplings}
\label{rencoup}

The renormalized couplings are calculated using Eq.~(\ref{eq:JWK}) or
(\ref{eq:J_K}) without and with the Korringa relaxation for different
bias voltages. In Fig.s~\ref{fig3a} and \ref{fig3b} the couplings
calculated from the sharp double-step distribution function given by
Eq.~(\ref{eq:freesol}) are shown together with the corresponding
distribution functions. Without Korringa relaxation the resonances
formed in the regions of the steps are very pronounced. Their widths
can be interpreted as the renormalized Kondo temperature. For finite
biases they are reduced due to the smaller steps with sizes $x$ or
$1-x$ in the distribution function. The larger the bias the narrower
the resonances are. At the overlap of the resonances the coupling can
be essentially enhanced.

\begin{figure}
 \begin{picture}
 (160,540)(10,10)
 \put(-120,-30){\includegraphics[height=20cm]{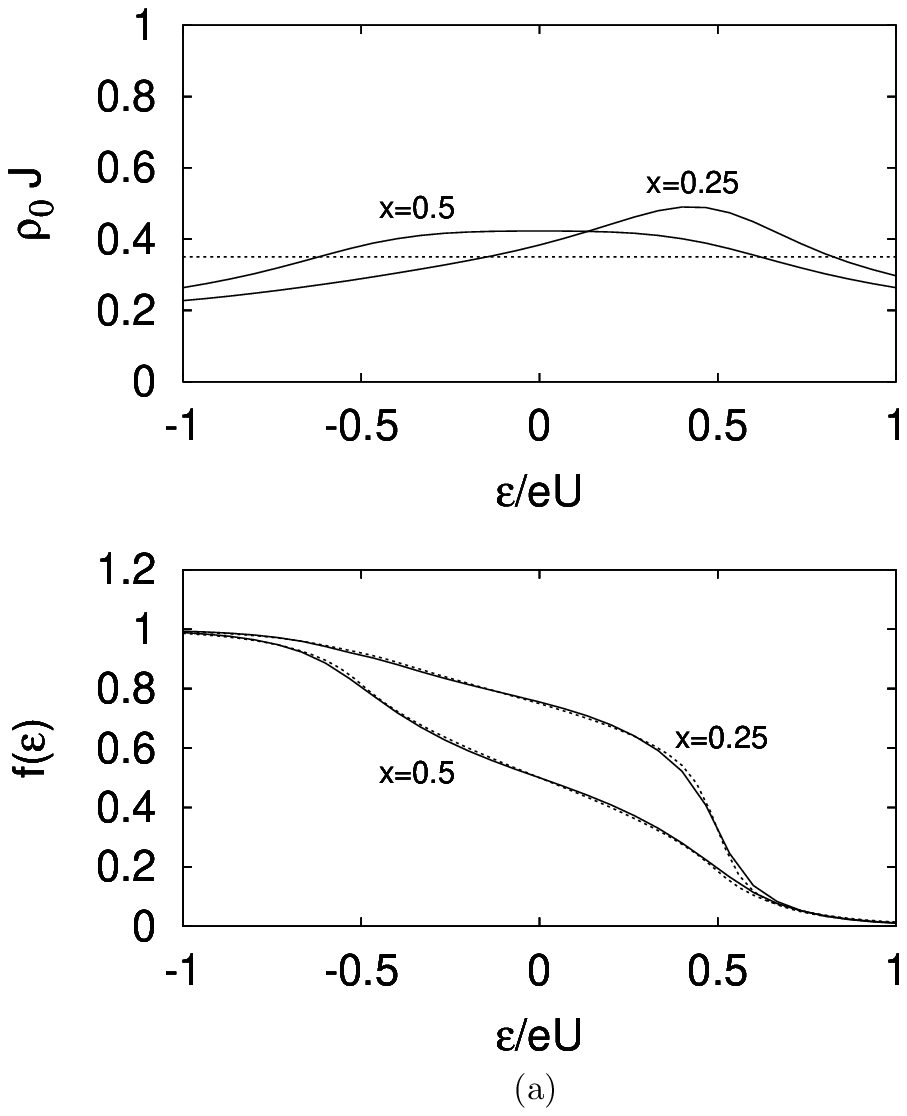}}
 \put(-120,-260){\includegraphics[height=20cm]{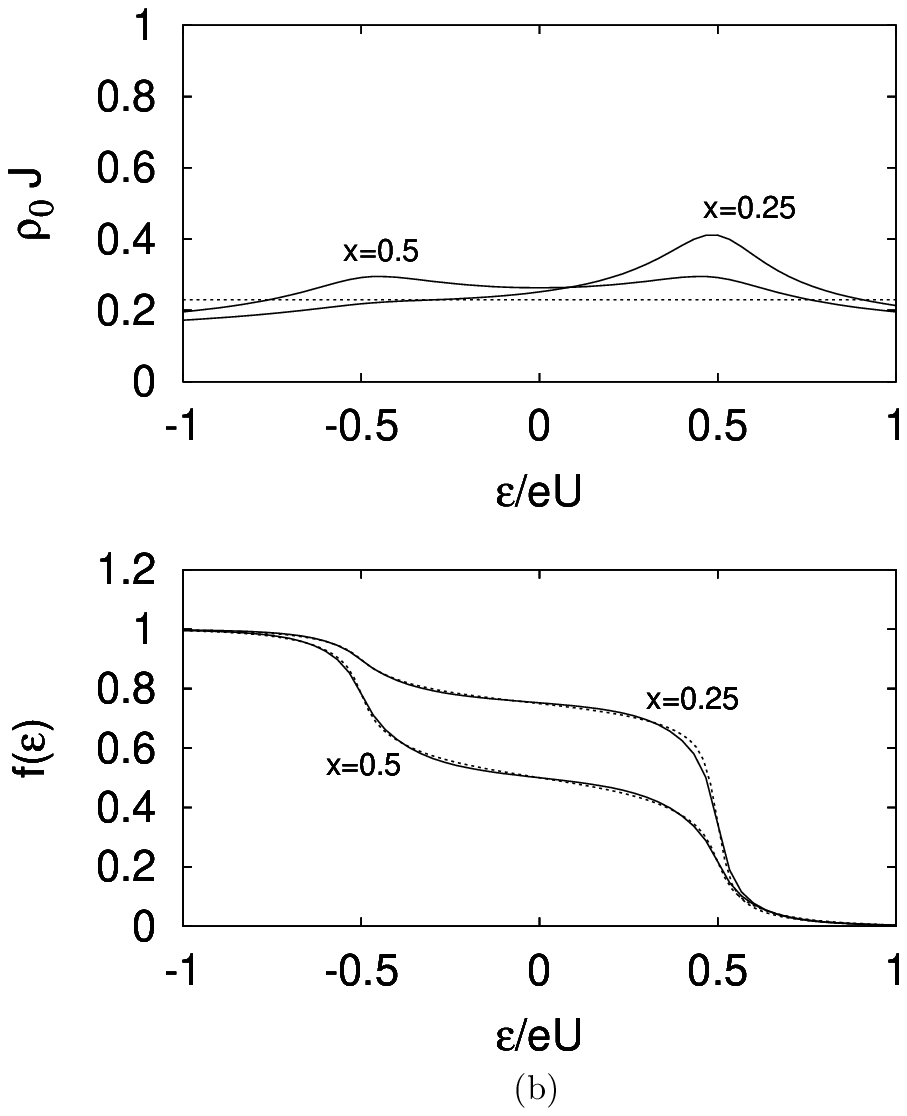}}
 \end{picture}
\caption{The coupling constant (top) and the distribution function (bottom) 
  obtained from the self-consistent calculation using the leading logarithmic
  renormalized coupling (solid line) and an appropriate constant coupling
  $\rho_0 {\bar J}$ (dotted line). (a) $U=0.15$mV, $\rho_0 {\bar J}=0.35$, (b)
  $U=0.3$mV, $\rho_0 {\bar J}=0.23$. The other parameters are $\tau_D=1.8$ns, 
  $c=4$ppm, $T_K=0.15$K, $\rho_0=0.21$/(site eV), $S=1/2$, $T=0.05$K.}
\label{fig6}
\end{figure}

\begin{figure}
\centerline{\includegraphics[height=5cm]{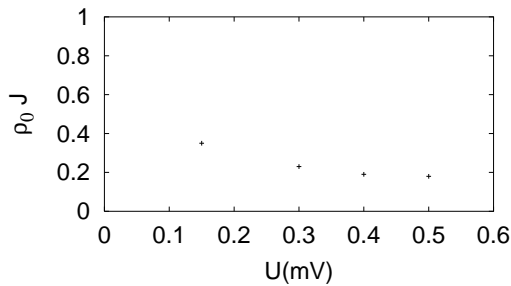}}
\caption{The dependence of the averaged, approximating coupling 
  $\rho_0 {\bar J}$ on the voltage bias for $T_K=0.15$K. The other parameters
  are $\tau_D=1.8$ns, $c=4$ppm, $\rho_0=0.21$/(site eV), $T=0.05$K.}
\label{fig_Udep}
\end{figure}

In Fig.~\ref{fig3a}(b) and (c) (Fig.~\ref{fig3b}(b) and (c)) just for
demonstration a fixed value of the Korringa relaxation
$\frac{\hbar}{2\tau_K}=0.25$K ($\frac{\hbar}{2\tau_K}=0.4$K) is used.
In reality the effect of the Korringa broadening is larger for larger
bias as $\tau_K^{-1}$ is linear in $U$.  As it can be seen in
Fig.s~\ref{fig3a}(c) and \ref{fig3b}(c) with $U=0.3$mV the peaks are
almost completely smeared.  In that case the coupling can be well
approximated by a constant value $\rho_0 {\bar J}$, but it is enhanced
compared to the bare one, e.g. in Fig.~\ref{fig6}(a) and (b) ${\bar
  J}/J_0\sim 9$ and ${\bar J}/J_0\sim 5.5$, respectively. In that case
the Korringa broadening can be estimated according to
Eq.~(\ref{Korr_eq}). For $S=1/2$, $x=0.5$, $e U=0.15$meV and $\rho_0
{\bar J}=0.35$ ($e U=0.3$meV and $\rho_0 {\bar J}=0.23$) the
relaxation rate is $\frac{\hbar}{2\tau_K}=0.22$K
($\frac{\hbar}{2\tau_K}=0.19$K). For larger bias the dependence of the
approximating constant coupling on the voltage is negligible (see
Fig.~\ref{fig_Udep}). (That is the reason why the same values of
$\frac{\hbar}{2\tau_K}$ were chosen in Fig.~\ref{fig6}(a) and (b) for
the different biases $e U=0.15$meV and $e U=0.3$meV.) In this way
the Korringa broadening can exceed the width of the resonance reduced
by the bias and that can be so effective that the resonances disappear
and the effect of the bias is only in the averaged coupling $\rho_0
{\bar J}$. In this way the dependence of the final result on
$\frac{\hbar}{2\tau_K}$ is very weak.

In Fig.~\ref{fig6} we show that the solution of the self-consistent
calculation using the leading order renormalized coupling (solid line)
can be well approximated by a constant coupling (dotted line).  The
parameters used in the calculation are $\tau_D=1.8$ns, $c=4$ppm,
$\rho_0=0.21$/(site eV), $S=1/2$, $T=0.05$K, $T_K=0.15$K. 

Based on that observation, the approximation in which the
self-consistent calculation is performed by using an appropriate
constant value for the coupling is applied in Ref.~[\onlinecite{UJZ}]
and Paper II where the surface magnetic anisotropy for the impurity
spin is included.

It is important to note, however, that the approximating constant coupling
depends on the voltage bias and on the bare coupling ($\sim T_K$).  The
former is illustrated in Fig.~\ref{fig_Udep} for $T_K=0.15$K.

\begin{figure}
\centerline{\includegraphics[height=4.3cm]{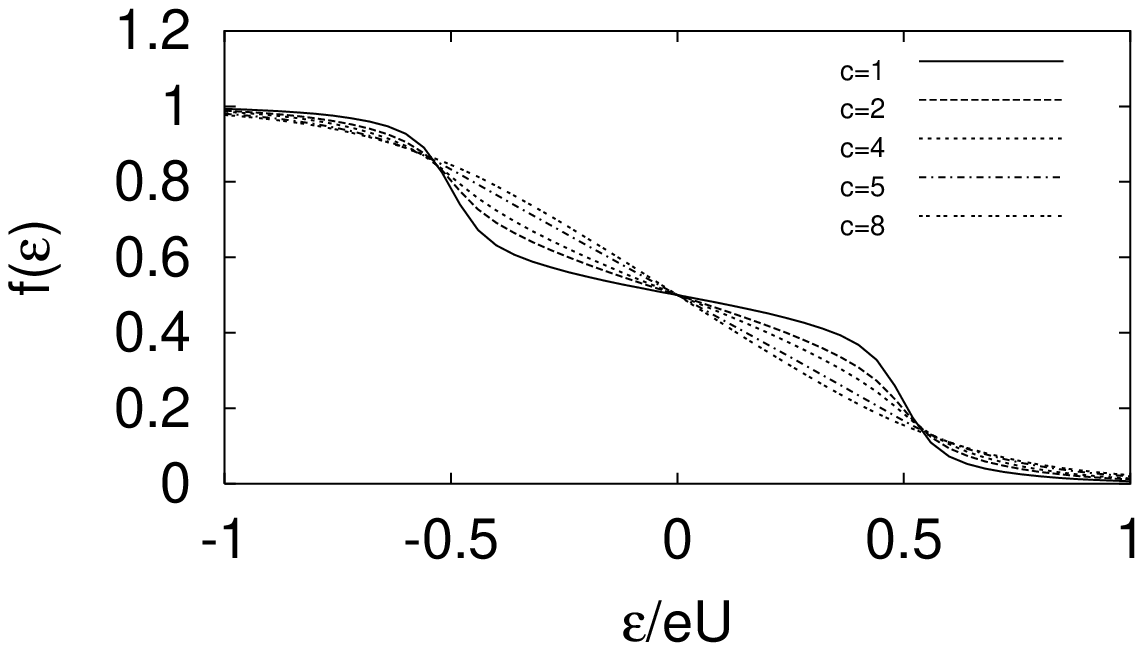}}
\centerline{(a)}
\vspace*{-2.7em}
\centerline{\includegraphics[height=4.3cm]{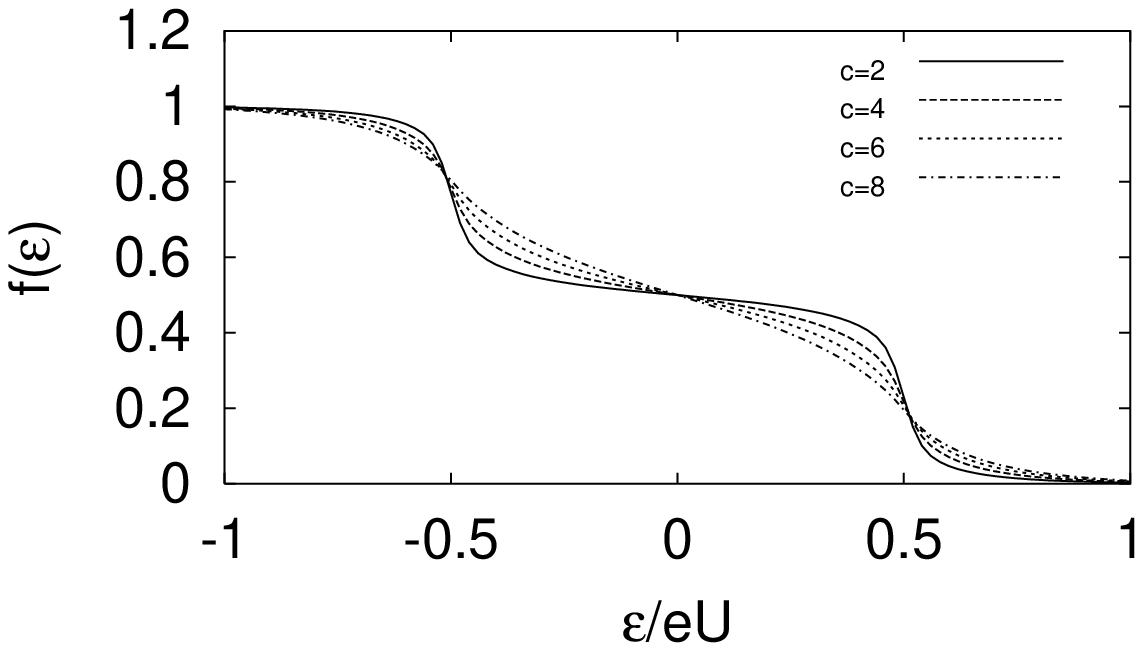}}
\centerline{(b)}
\caption{The dependence of the distribution functions on the impurity
  concentration is illustrated using the appropriate constant
  couplings obtained (see Fig.~\ref{fig6}). (a) $U=0.15$mV, $\rho_0 {\bar
    J}=0.35$, (b) $U=0.3$mV, $\rho_0 {\bar J}=0.23$.  The other
  parameters are $\tau_D=1.8$ns, $T_K=0.15$K, $\rho_0=0.21$/(site eV),
   $T=0.05$K.}
\label{fig_cdep}
\end{figure} 
The dependence of the distribution function on the impurity
concentration is shown in Fig.~\ref{fig_cdep} for (a) $U=0.15$mV and
(b) $U=0.3$mV using the appropriate constant couplings obtained from
Fig.~\ref{fig6}. The concentration dependent correction is linear only
for lower concentrations.

\subsection{Distribution function}

The calculations are performed according to the iteration scheme described
above. In case of $\tau_K=\infty$ the kernel given by Eq.~(\ref{runkernel}) is
singular in the energy transfer $E$ as $E^{-2}$. There is, however, the cross
term with the less singular denominator $E (\ep-\ep'-E)$. That term is less
important because of the weaker singularity. Furthermore that drops out when
the energy dependence of the couplings is neglected. 

\begin{figure}
 \begin{picture}
 (90,230)(10,50)
 \put(-120,-150){\includegraphics[height=18cm]{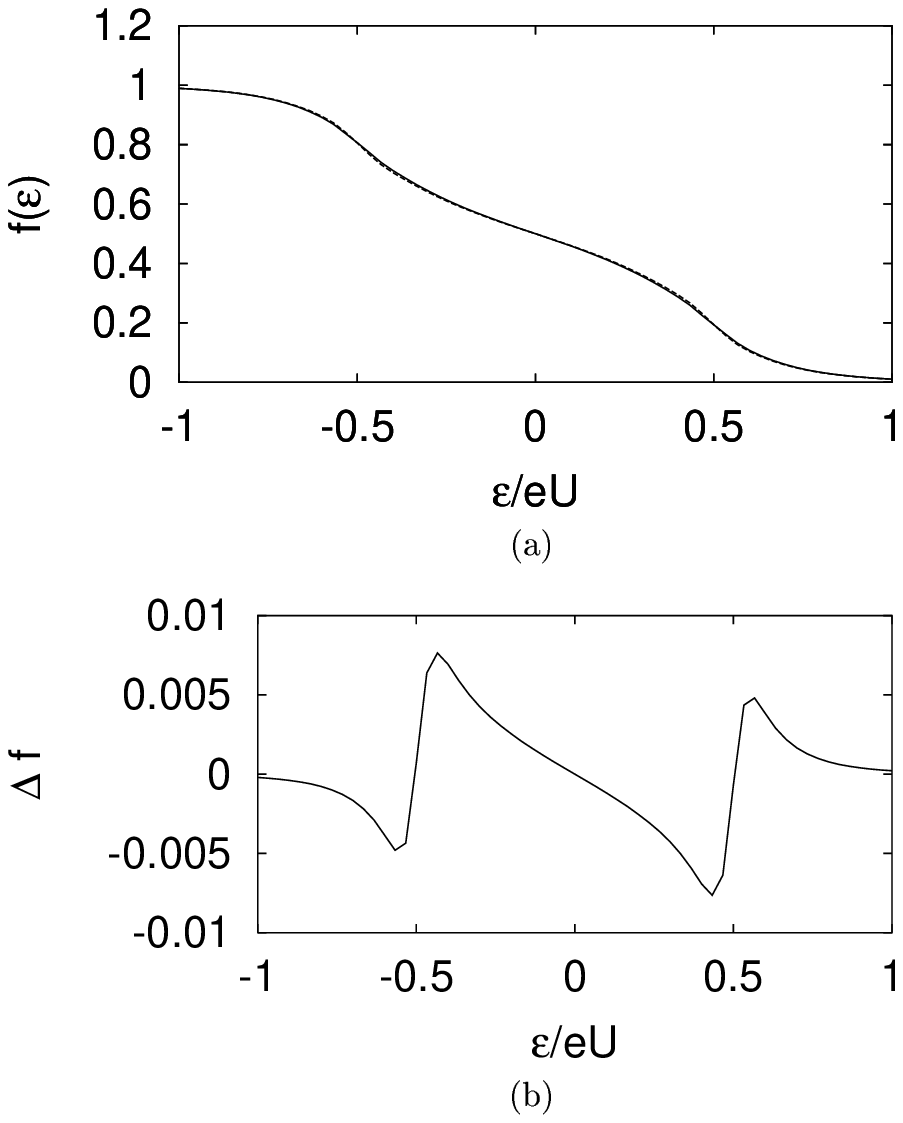}}
 \end{picture}
\caption{Comparison of the solutions of the self-consistent calculation 
  with and without the cross terms in the kernel Eq.~(\ref{korrkernel}). (a)
  the distribution functions with (solid line) and without the cross terms
  (dotted line) (b) the difference between the two solutions. $\tau_D=1.8$ns,
  $c=4$ppm, $\rho_0=0.21$/(site eV), $S=1/2$, $x=0.5$, $T=0.05$K, $T_K=0.3$K,
  $U=0.3$mV.}
\label{fig4}
\end{figure}

In the presence of finite Korringa relaxation the situation is more complex
(see the kernel given by Eq.~(\ref{korrkernel})). The cross term does not
disappear even for constant coupling, but it has limited importance as it is
demonstrated by calculating the final distribution function without and with
the cross term.  In Fig.~\ref{fig4} the role of the cross terms in the kernel
given by Eq.~(\ref{korrkernel}) is illustrated (a) where the solution of the
self-consistent calculation is plotted with (solid line) and without the cross
terms in the kernel (dotted line). The parameters used in the calculation are
$\tau_D=1.8$ns, $c=4$ppm, $\rho_0=0.21$/(site eV), $S=1/2$, $T=0.05$K,
$T_K=0.3$K, $U=0.3$mV, and $x=0.5$. The difference is minor which can be seen
from Fig.~\ref{fig4}(b). Deviation occurs only in the vicinity of the steps.
Thus in most of the cases neglection of the cross terms is a good
approximation.

\subsection{The effect of the ambiguity in the strength of the Korringa
 relaxation} 

\begin{figure}
\centerline{\includegraphics[height=4.3cm]{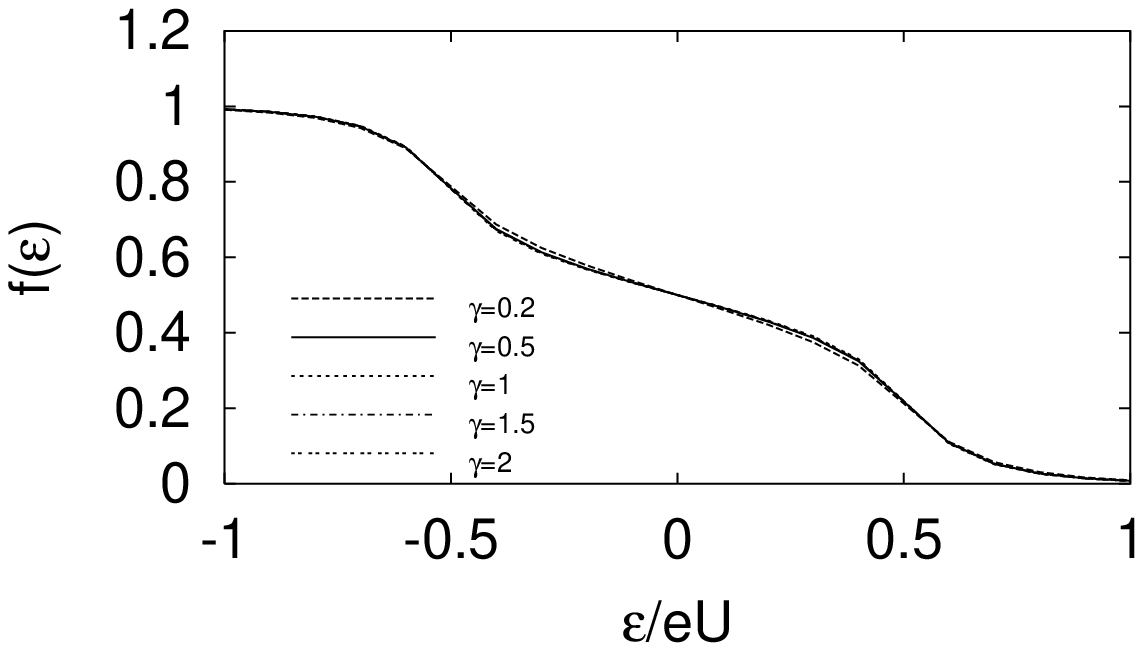}}
\centerline{(a)}
\vspace*{-2.7em}
\centerline{\includegraphics[height=4.3cm]{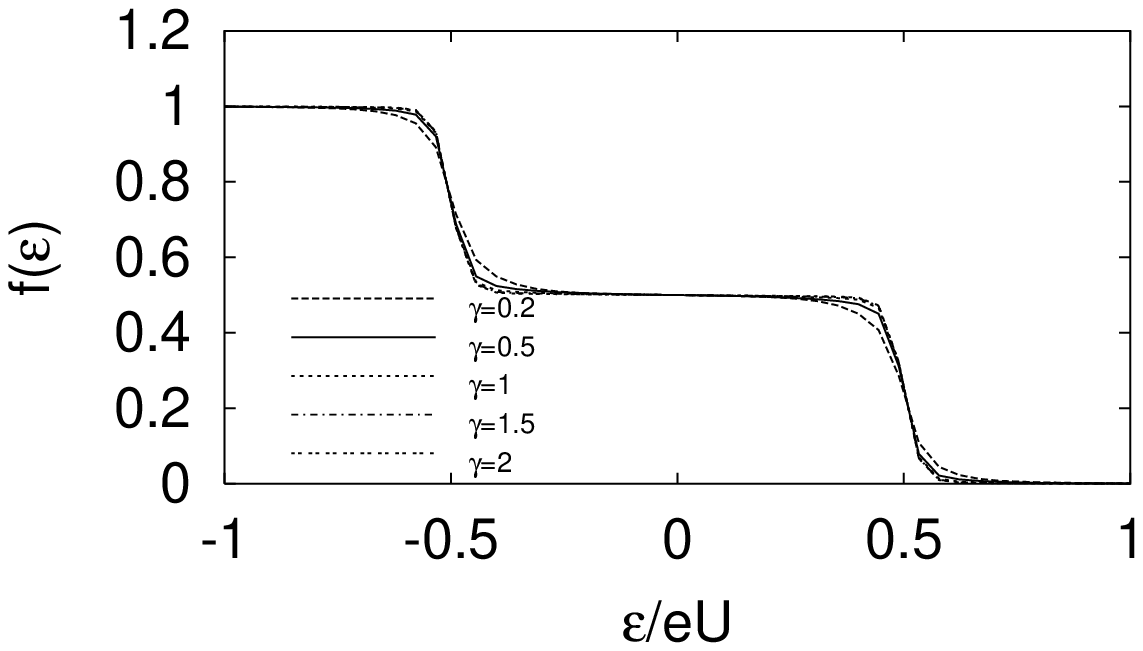}}
\centerline{(b)}
\caption{The distribution functions are illustrated where the Korringa
  relaxation rates are modified by a multiplying factor; $\gamma=\lambda S
  (S+1)$. (a) $U=0.1$mV (b) $U=0.3$mV. The other parameters are
  $\tau_D=2.8$ns, $c=5$ppm, $\rho_0=0.21$/(site eV), $T_K=0.3$K, $T=0.05$K.}
\label{fig5}
\end{figure}

It has been pointed out earlier that the anomalously strong Korringa
relaxation in the non-equilibrium case is not treated in a consequent way as
the vertex corrections and the broadening of the initial and final states are
not taken into account properly. In order to judge their importance a
multiplicative factor $\lambda$ is introduced as
$\frac{\hbar}{2\tau_K}\longrightarrow \lambda\frac{\hbar}{2\tau_K}$ (later the
notation $\gamma=\lambda S (S+1)$ is used).  Two sets of curves are calculated
for biases $0.1$mV and $0.3$mV (see Fig.~\ref{fig5}). In both cases $\gamma$
is changed in the wide interval $0.2\leq\gamma\leq 2$. The differences are
relatively very small as $\gamma$ exceeds the value $\gamma\sim 0.4$. That is
a consequence of the large smearing in the coupling and even more in the
distribution function. Thus, the improper treatment of the vertex corrections
is not expected to show up as drastic changes in the results.

\subsection{Distribution function at points out of the middle}

\begin{figure}
\centerline{\includegraphics[height=5cm]{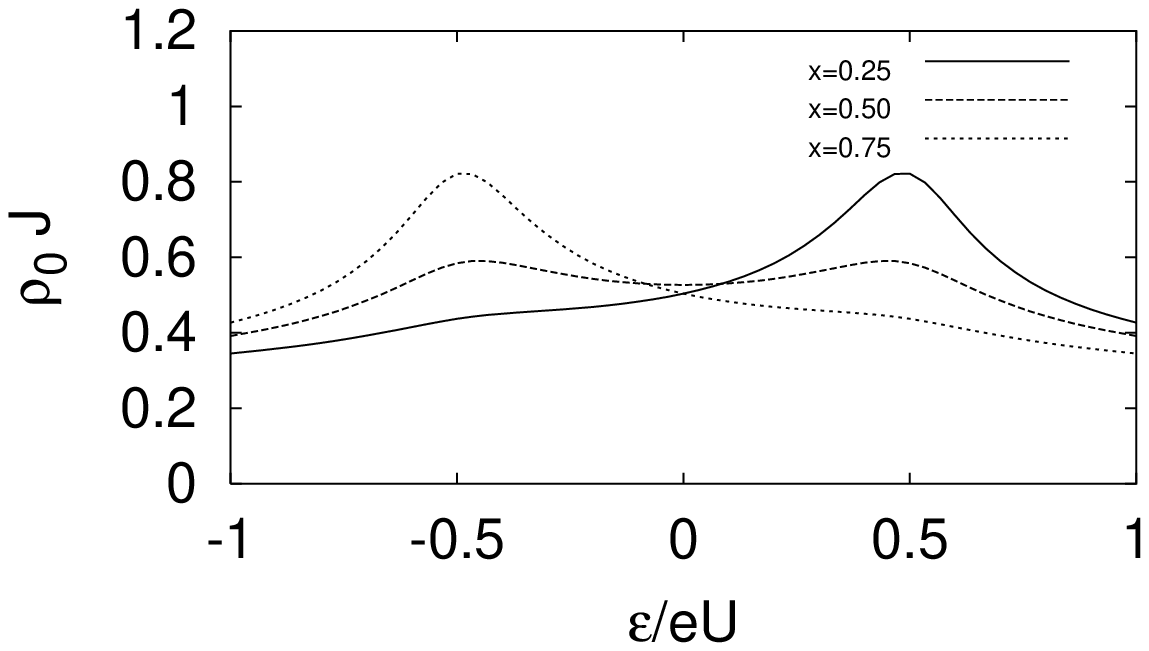}}
\vspace*{-2em}
\centerline{\includegraphics[height=5cm]{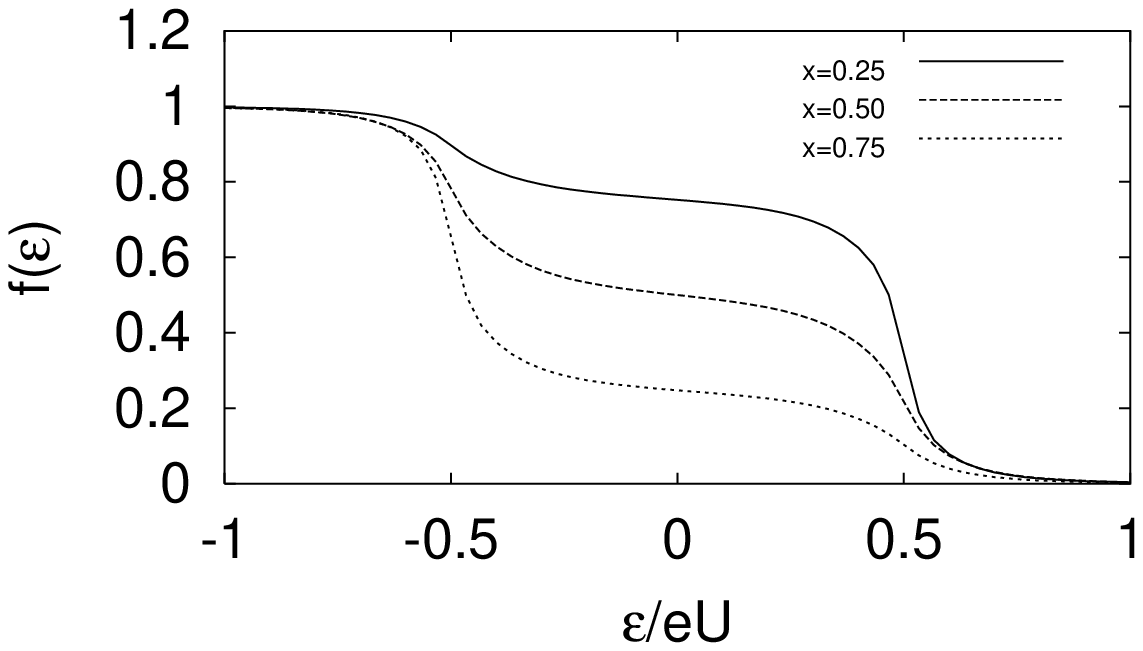}}
\caption{The coupling constant (top) and the distribution function (bottom) 
  obtained from the self-consistent calculation at different positions
  in the wire.  The other parameters are $U=0.3$mV, $\tau_D=1.8$ns,
  $c=4$ppm, $\rho_0=0.21$/(site eV), $T_K=0.15$K, $T=0.05$K.}
\label{fig_x}
\end{figure} 
The calculation can be performed at any position in the wire and the
results are demonstrated in Fig.~\ref{fig_x}. In these cases the
reduction in the widths of the Kondo resonances are different at the
two steps, namely at the smaller step it is more reduced. In the
distribution function the height of the steps is asymmetric according
to the actual value of $x$. 

\subsection{Scaling in terms of the applied voltage bias}

Already in the early experiments \cite{Pothier} the scaling with respect to
the applied voltage has been observed, thus the distribution function at a
given position $x$ was described by a single variable function
$f_x(\frac{\ep}{e\,U})$. Since that time the validity of the scaling has
played a central issue both in experimental \cite{Pothier2} and theoretical
studies \cite{KG,th1,th4}. 

\begin{figure}
 \begin{picture}
 (160,500)(10,10)
 \put(-140,-40){\includegraphics[height=22cm]{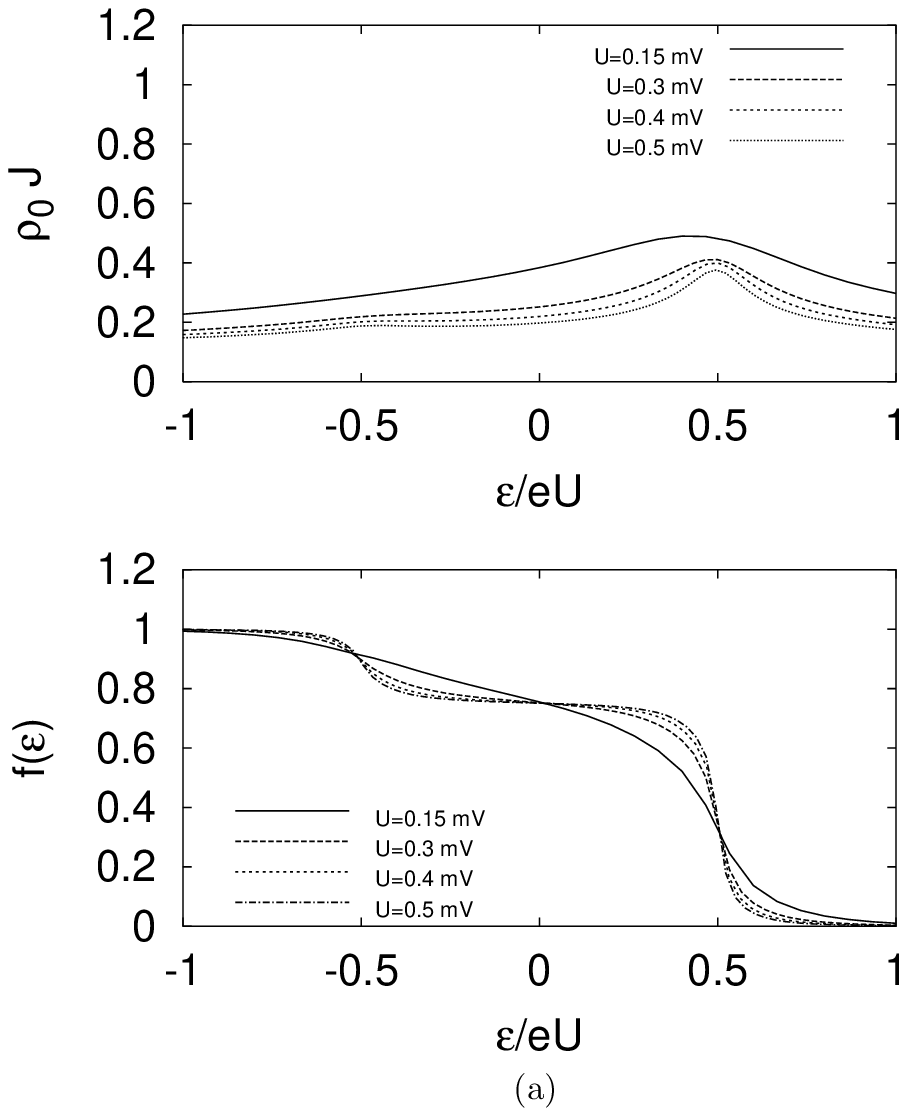}}
 \put(-140,-290){\includegraphics[height=22cm]{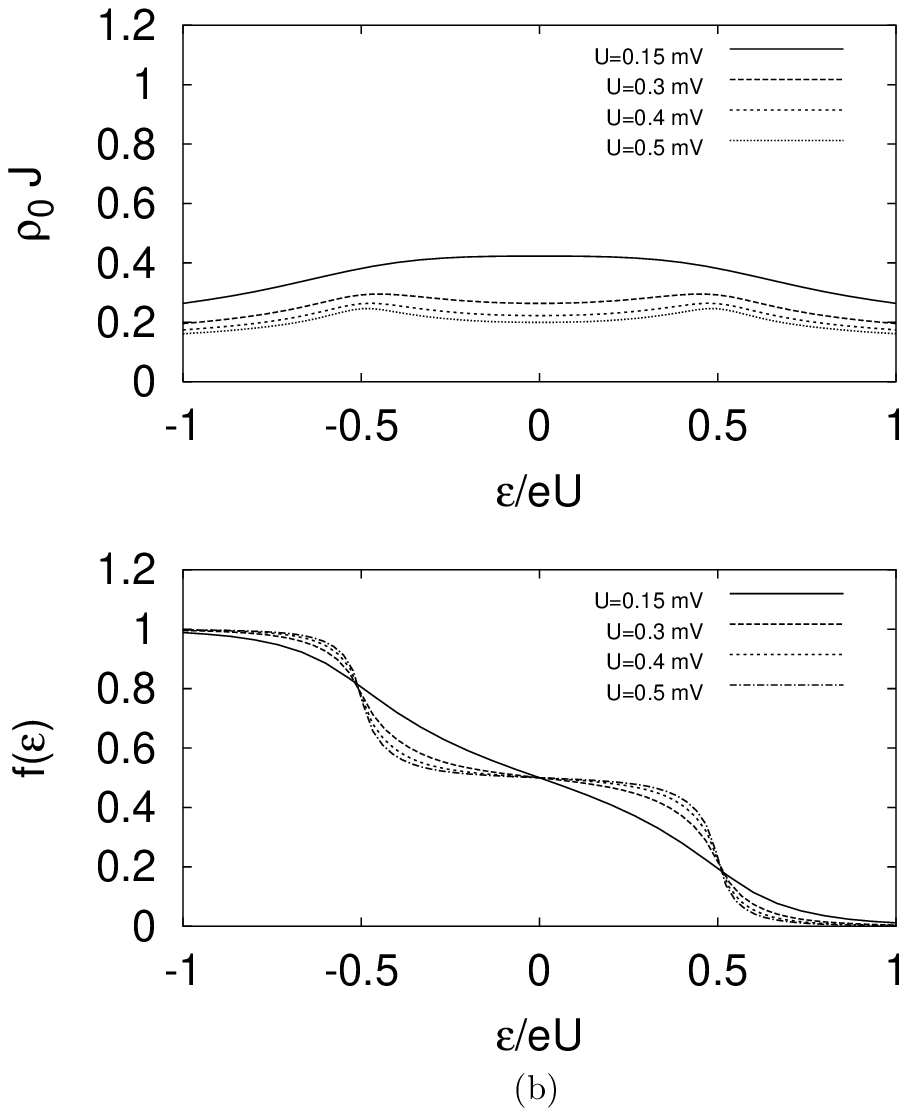}}
 \end{picture}
\caption{The coupling constant (top) and the distribution function (bottom) 
  obtained from the self-consistent calculation for different biases. (a)
  $x=0.25$, (b) $x=0.5$. The other parameters are $\tau_D=1.8$ns,
  $c=4$ppm, $T_K=0.15$K, $\rho_0=0.21$/(site eV), $T=0.05$K.}
\label{fig_scaling}
\end{figure}

In the present calculation the distribution function at position $x$
is a function of the energy $\ep$ and voltage bias $U$ and it is a
functional of the self-consistently determined coupling
$J(\ep,e\,U;x)$, thus it can be written as $f_{\{J\}}(\ep,e\,U;x)$. It
has been demonstrated in Section~\ref{rencoup} that the coupling
$J(\ep,e\,U;x)$ is a slowly varying function of the energy for a fixed
value $e\,U$ in the relevant energy window with width $e\,U$, which is
the dominant one concerning the non-equilibrium Boltzmann equation
(\ref{BE}) case. Thus in case of slowly varying $J$ shown in
Fig.s~\ref{fig3a}, \ref{fig3b} and \ref{fig6} the coupling
$J(\ep,x)$ can be approximated by an averaged one, ${\bar J}(e\,U)$
which is independent of $x$ as it is also illustrated in
Fig.~\ref{fig6}. In that case the distribution function is simplified
as $f_{\{J\}}(\ep,e\,U;x) \Rightarrow f(\frac{\ep}{e\,U},{\bar
  J(e\,U)};x)$.  Now, if $J$ only slightly depends on the voltage bias
$U$ (see Fig.~\ref{fig_Udep} for $T_K=0.15$K) then the bias dependence
can be dropped for a limited interval of $e\,U$ and ${\bar J}=const.$
can be used.  Then the distribution function is $f_{\{J\}}(\ep,e\,U;x)
\Rightarrow f(\frac{\ep}{e\,U},{\bar J};x)$ which exhibits the scaling
observed experimentally in several cases and illustrated in
Fig.~\ref{fig_scaling}.



\section{Conclusion}
\label{sec:6}

In this paper the non-equilibrium transport is studied in the presence of
magnetic impurities representing the 1-channel Kondo (1CK) problem.  The
results can be generalized to the 2CK problem which may be realized by
dynamical structural defects \cite{structdef}. In the logarithmic
approximation scheme applied in the present paper there is no difference in
the leading logarithmic order as the channel index silently follows the
continuous electron line. The first differences occur in the next
approximation as closed electron loops appear where summation is required with
respect to the channel index, thus the Korringa relaxation and the vertex
corrections are multiplied by the number of channels. As it has been discussed
previously no drastic effects can be expected.

There is an other drawback of the application of the logarithmic
approximation in the leading order as that overestimates the Kondo
temperature by missing the prefactor $(2\rho_0 J)^{1/2}$. In general
the contribution of the next to leading logarithmic approximation can
be essential, but only nearby the Kondo temperature. The Korringa
broadening reduces also that effect by reducing the averaged coupling
$\rho_0 {\bar J}$. On the other hand if the Kondo temperature is comparable
with the applied voltage, then the effective Kondo temperature is less
reduced, thus the present method cannot be applied \cite{th4,Coleman}.

In the present scheme the kernel of the effective magnetic impurity
induced electron-electron interaction is determined. That kernel is
governed by the singularity $E^{-2}$ in the energy transfer and the
less singular cross term results only in minor changes. For determination of
the distribution function a completely self-consistent calculation is
carried out where in the framework of the ``poor man's'' version of the
renormalization group the couplings are modified by reducing the bandwidth 
which also modifies the Korringa rate and the electron
distribution function.

For an arbitrary set of the parameters the distribution function can be
determined using the method developed for the weak coupling limit, where $T_K$
is small compared to the applied voltage . The relative importance of the
different ingredients can be judged as the method is applied step by step.  In
the present formulation some details could be easier followed, than e.g. in
the non-crossing approximation which is more powerful but many details are
buried by the huge machinery.

Considering the comparison with the experiments an essential dilemma
has been earlier realized. The required impurity concentration to fit
the experiments in case of spin $S=5/2$ magnetic impurities is in
good accordance with values obtained by the other experiments
determining the dephasing time. In contrary, the cases with presumable
integer spins show drastic discrepancies. The dephasing requires
sometimes less than one hundred of the impurity concentration and that
was attributed to the surface anisotropy \cite{UZ1} in
Ref.~\onlinecite{UJZ}. 

The revised version of the form of the surface anisotropy \cite{UZ1}
makes the effect on the electron distribution much weaker, due to its
smaller amplitude at larger distances measured from the surface. As in
Ref.~\onlinecite{UJZ} it is shown that the surface anisotropy has weak
effect on the shape of the distribution, thus the reduced number of
impurities with large anisotropy very likely cannot change the
distribution in an essential way. 

Concidering the dephasing rate, that effect may, however, be stronger,
as the relevant energy range of the electrons is much smaller, as the
applied voltage $eU$ is replaced by the temperature ($eU>kT$). Thus,
for more impurities, the anisotropy can dominate over the thermal
energy, and the dephasing rate can be more reduced.

\section*{ACKNOWLEDGMENTS}
We are grateful to N. O Birge, V. M. Fomin, H. Grabert, J. Kroha, H.
Pothier, P. W\"olfle, and G. Zar\'and for useful discussions.  This
work was supported by Hungarian grants OTKA F043465, T046303, T048782,
TS040878, T038162, grant No. RTN2-2001-00440, and MTA-NSF-OTKA Grant
INT-0130446. One of us (A.Z.) thanks N. Andrei and P. Coleman for
comments and the kind hospitality at Rutgers University.

\appendix*
\section{}

The derivation of the result Eq.~(\ref{runkernel}) is
outlined in the non-vanishing leading logarithmic order. The Abrikosov's
pseudofermion technique \cite{Abrikosov} is applied where the $b_{\alpha}$
operators create spin states with index $\alpha$ and then ${\bf
  S}=\sum\limits_{\alpha\beta} b^{\dagger}_{\alpha} {\bf S}_{\alpha\beta}
b_{\beta}$. For simplicity the time ordered diagrams are used with real
time. That technique can be applied starting with arbitrary states
(e.g. non-equilibrium) and the transition amplitude between states 
 $|i\rangle$ and $\langle f|$ is given as
\begin{equation}
  \langle f|T|i\rangle_\omega = \sum_{n=1}^\infty \langle f|H_1
  \left(\frac 1 {\omega+i\delta -H_0} H_1 \right)^n |i\rangle,
\label{eq:fTi}
\end{equation}
where $\omega$ is the energy of the initial state.

The scattering rate can be calculated in two different ways:
(i) The scattering amplitudes $\Gamma$ are evaluated using Eq.~(\ref{eq:fTi})
and then those are inserted in the golden rule. (ii) The imaginary part of the
scattering amplitude is combined with the optical theorem (that method is used
in Ref.~\onlinecite{th3}). 

\begin{figure}
\centerline{\includegraphics[height=2.5cm]{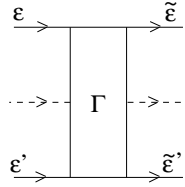}}
\caption{The two electron and one pseudofermion scattering amplitude
  $\Gamma$. The solid and dotted lines represent the electrons and the
  pseudofermion, respectively. One of the electron lines connects continuously
  the upper corners in the amplitude $\Gamma$ and the other the lower corners.}
\label{Afig1}
\end{figure}

In the first method the direction of the spin lines coincides with the time
evolution as only one spin state can be present at a given time. In the second
method the two electron scattering amplitude is determined like in the
original work of S\'olyom and Zawadowski\cite{SZ} and the spin line is closed.

\begin{figure}
\centerline{\includegraphics[height=7cm]{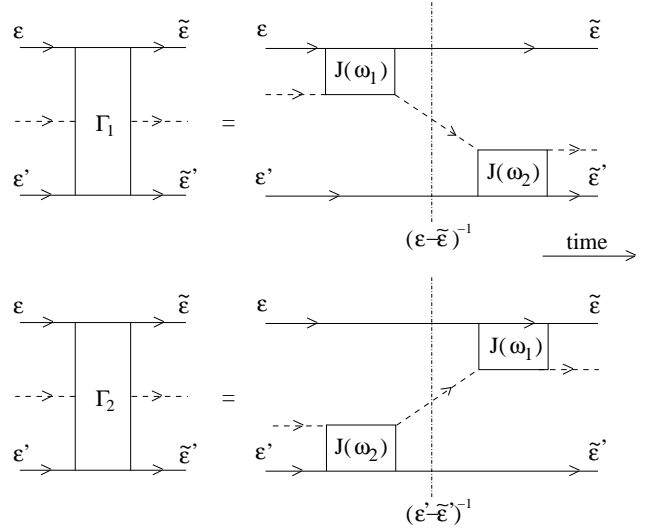}}
\caption{The two basic diagrams providing singular terms in the energy 
  transfer. The solid and dotted lines represent the electrons and the
  pseudofermion, respectively. The square box represents the logarithmic
  electron-pseudofermion vertex correction $J(\omega)$ and the dotted line
  the position of the energy denominator.}
\label{Afig2}
\end{figure}

In the following the first method is used like in the papers by Kaminski and
Glazman\cite{KG} and G\"oppert and Grabert\cite{th1}. The two basic diagrams
providing singular terms in the energy transfer $E=\ep-\tilde{\ep}$ can be
seen in Fig.~\ref{Afig2} where the logarithmic vertex corrections $J(\omega)$
with two electron and two pseudofermion legs are represented by square boxes.
In these amplitudes the energy is conserved, thus
$\ep+\ep'=\tilde{\ep}+\tilde{\ep}'$. The energy denominators indicated by
dotted lines in Fig.~\ref{Afig2} separate the two vertex corrections
$\Gamma_1$ and $\Gamma_2$ and they are $\ep-\tilde{\ep}=E$ and
$\ep'-\tilde{\ep}'=\tilde{\ep}-\ep=-E$ where in the latter the energy
conservation is used. These are singular in the energy transfer $E$.

As in case of absence of magnetic field and surface anisotropy the spins do
not carry energy and the incoming and outgoing electrons are on the energy
shell, thus the energy variables of the vertices are $\omega_1=\ep$,
$\omega_2=\tilde{\ep}'=\ep+\ep'-\tilde{\ep}=\ep'+E$ in $\Gamma_1$
and $\omega_2=\ep'$, $\omega_1=\tilde{\ep}=\ep-E$ in $\Gamma_2$.
In terms of the energy dependent vertex $J(\omega)$ which is real in the
logarithmic approximation the contributions are proportional to $J(\ep)
J(\ep'+E)$ and $J(\ep') J(\ep-E)$, respectively.

\begin{figure}
\includegraphics[width=8.5cm]{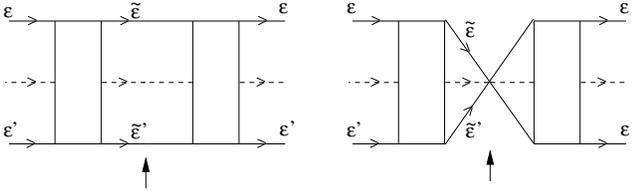}
\caption{The two forward scattering diagrams coming in the golden rule. 
  The solid and dotted lines represent the electrons and the
  pseudofermion, respectively. The square boxes represents the appropriate
  vertexes ($\Gamma_1$ or $\Gamma_2$) and the arrows show from where the
  imaginary parts are taken.}
\label{Afig3}
\end{figure}
Applying the golden rule the sum of the contributions of these two diagrams
must be squared. In order to visualize the result of the detailed calculation
we show the two forward scattering diagrams in Fig.~\ref{Afig3} in the spirit
of the optical theorem where the arrows indicate from where the imaginary
parts are taken, thus the energy is conserved.  Considering the second
``crossing'' diagram in the vertex on the right hand side the energy variables
of the final states $\tilde{\ep}\leftrightarrow \tilde{\ep}'$ are interchanged
in comparison to the first diagram (see Fig.~\ref{Afig3}).  That results also
in changing the energy denominators like $\ep-\tilde{\ep}=E \rightarrow
\ep-\tilde{\ep}'=-\ep'+\tilde{\ep}=\ep-\ep'-E$ and $\ep'-\tilde{\ep}'=-E
\rightarrow \ep'-\tilde{\ep}=-(\ep-\ep'-E)$ and also
$\omega_2=\ep'+E\rightarrow \omega_2=\tilde{\ep}=\ep-E$ in $\Gamma_1$ and
$\omega_1=\ep-E\rightarrow \omega_1=\ep+\ep'-\tilde{\ep}= \ep'+E$ in
$\Gamma_2$.

In the ``non-crossing'' term the energy denominators lead to a $1/E^2$
singularity. The vertex corrections corresponding to the terms proportional to
$\Gamma_1^2$, $\Gamma_2^2$ and $\Gamma_1 \Gamma_2$ ($\Gamma_2 \Gamma_1$) are 
$J(\ep)^2 J(\ep'+E)^2$, $J(\ep')^2 J(\ep-E)^2$ and $J(\ep) J(\ep'+E) J(\ep')
J(\ep-E)$ occur in Eq.~(\ref{runkernel}). 

In the ``crossing'' contribution, however, the singularity has the form
$\bigl (E(\ep-\ep'-E)\bigr )^{-1}$. The corresponding vertex corrections
are $J(\ep) J(\ep'+E) J(\ep) J(\ep-E)$, $J(\ep) J(\ep'+E) J(\ep')
J(\ep'+E)$, $J(\ep') J(\ep-E) J(\ep) J(\ep-E)$, $J(\ep') J(\ep-E) J(\ep')
J(\ep'+E)$ in the terms $\Gamma_1^2$, $\Gamma_1 \Gamma_2$, $\Gamma_2 \Gamma_1$,
and $\Gamma_2^2$. The vertex corrections with these energy variables occur in
Eq.~(\ref{runkernel}), as well.

\end{document}